\documentclass[pra,aps,10pt,twocolumn,amsmath,amssymb]{revtex4-1}

\usepackage[english]{babel}
\usepackage{afterpage}
\usepackage{graphicx}% Include figure files
\usepackage{dcolumn}% Align table columns on decimal point
\usepackage{bm}% bold math
\usepackage{color}
\usepackage{slashed}%slashed D, k, A, etc.
\usepackage{latexsym}
\usepackage{appendix}
\usepackage{esint}
\usepackage{physics}
\usepackage{epstopdf}
\usepackage{mathtools}

\newcommand{\ii}{\mathrm{i}} %complex i
\newcommand{\ee}{\mathrm{e}} %eksponent
\newcommand{\me}{m_\mathrm{e}} %electron mass

\AppendGraphicsExtensions{.gif}

\begin{document}

%\title{ Creation and control of electron vortices in photodetachment: The role of the carrier-envelope phase}
\title{Carrier-envelope-phase and helicity control of electron vortices in photodetachment}
\author{M. M. Majczak}
\author{F. Cajiao V\'elez}
\email{Felipe.Cajiao-Velez@fuw.edu.pl}
\author{J. Z. Kami\'nski}
\author{K. Krajewska}
\email{Katarzyna.Krajewska@fuw.edu.pl}

\affiliation{
Institute of Theoretical Physics, Faculty of Physics, University of Warsaw, Pasteura 5, 02-093 Warsaw, Poland}
\date{\today}

\begin{abstract}
Formation of electron vortices and momentum spirals in photodetachment of the H$^-$ anion driven by isolated ultrashort
laser pulses of circular polarization or by pairs of such pulses (of either corotating or counterrotating polarizations) are analyzed under the scope of the strong-field approximation. 
It is demonstrated that the carrier-envelope phase (CEP) and helicity of each individual pulse can be used to actively manipulate and control 
the vortical pattern in the probability amplitude of photodetachment. Specifically, the two-dimensional mappings of probability amplitude
can be rotated in the polarization plane with changing the CEP of the driving pulse (or two corotating pulses); thus,
offering a new tool of field characterization. Furthermore, it is shown that the formation of spirals or 
annihilation of vortices relates directly to the time-reversal symmetry of the laser field, which is realized by a pair of 
pulses with opposite helicities and CEPs.
\end{abstract}

\maketitle

\section{Introduction}
\label{sec:introduction}
During the last decade there has been an increasing interest towards the {\it electron momentum spirals} (often referred as `vortices') 
in laser-induced photoionization~\cite{Starace2015,Starace2016,Starace2017,Maxwell2020,Bandrauk2016,Bandrauk2017,Faria2021,Wollenhaupt2017,Wollenhaupt2019a,Wollenhaupt2019b,Wollenhaupt2020a,Wollenhaupt2020b} 
or photodetachment~\cite{Circular}. They manifest themselves, in the probability distribution of photoelectrons, as zones of large 
probability which follow concentric Fermat spirals with well-defined number of arms~\cite{Starace2015}. However, other type of structures, 
known as {\it electron vortices}, can also be found in the momentum distribution of photoelectrons; they appear as continuous lines in the 
three-dimensional momentum space where the probability amplitude of ionization (or detachment) vanishes, and its phase changes from zero 
to integer multiples of $2\pi$ around them~\cite{Dirac1931,Circular,VonKarmaan,GroundState,Trains}. Those two types of structures are 
fundamentally different as demonstrated by their physical properties; for instance, electron vortices carry a nonvanishing and quantized orbital-angular momentum (OAM)~\cite{Bliokh2017,Lloyd2017}, while spirals are characterized by a null OAM.

While electron vortices and spirals are formed due to the same physical phenomena, i.e., by subtile interference effects in the probability 
amplitude of ionization, whether one or the other are observed depends on the light field configuration and the target atom 
(or ion)~\cite{Maxwell2020,GroundState}. Several theoretical studies have predicted the formation of momentum spirals in photoionization 
from a variety of atomic and molecular targets~\cite{Starace2015,Bandrauk2016,Bandrauk2017,Faria2021} and diverse laser field 
configurations~\cite{Starace2016,Starace2017}. For instance, it has been shown that a sequence of two counterrotating circularly-polarized
and ultrashort laser pulses leads to momentum spirals, whereas single pulses or trains of corotating pulses lead to the formation of electron vortices~\cite{Circular}. However, corotating trains of bichromatic laser pulses may also lead to the spiral formation~\cite{Bandrauk2016,Bandrauk2017}. %However, for monochromatic trains of circularly polarized pulses, only the counterrotating configuration leads to spirals~\cite{Starace2015,Circular}. 

According to the analysis presented in Ref.~\cite{Starace2015}, which is performed in the perturbation regime of laser-matter interactions, the energy spectra 
of photoelectrons obtained in the photoionization of He atoms by two corotating laser pulses is circularly-symmetric; i.e., annular zones 
of zero probability, similar to Newton's rings, are observed. Moreover, a change of the relative carrier-envelope phase (CEP) between 
the driving pulses leaves the angular symmetry unchanged. Hence, the effect of the CEP on the electron distribution is, at most, detected 
as a variation of the rings radial locations or their intensities. Note that in the nonperturbative regime, the probability distribution 
of photoelectrons stimulated by corotating pulses is not circularly-symmetric, and vortical structures are observed together with the 
annular zones of zero probability~\cite{Circular}. 

In this paper, we further advance the theoretical understanding of vortical structures in photodetachment driven 
by circularly-polarized laser pulses. Various pulse configurations are considered, including isolated pulses and pairs of pulses in
corotating and counterrotating schemes. The calculations presented here are based on the strong-field approximation 
(SFA)~\cite{Keldysh,Faisal, Reiss} in a nonperturbative regime, which provides a remarkable agreement with ab initio methods 
of solving numerically the Schr\"odinger equation~\cite{Circular,VonKarmaan,Trains}. Our analysis focuses on the CEP effects over the formation of vortex structures. We demonstrate that
the vortical pattern in the probability amplitude of photoelectrons rotates in the polarization plane by the CEP of the driving pulse.
The same is observed in the corotating configuration of two identical pulses. For pulses with opposite helicities, on the other hand,
momentum spirals can be observed. More specifically, we show that the formation of spirals is closely related to annihilation of vortex-antivortex pairs, which occurs for 
laser fields with the time-reversal symmetry. Since the manipulation of electron vortical structures in photodetachment is sensitive 
to the handedness and the CEP of the laser field, it might provide additional means of field characterization.

We use atomic units (a.u.) along this paper. For our theoretical derivations we set $\hbar=1$ but show the electron charge, $e=-|e|$, 
and mass, $\me$, explicitly. Our numerical illustrations are presented in terms of the atomic units of momentum $p_{\rm at}=\alpha\me c$ 
and energy $E_{\rm at}=(\alpha c)^2\me$, where $\alpha$ is the fine-structure constant and $c$ is the speed of light. Furthermore,
the atomic unit of length corresponds to the Bohr radius, $a_0=\hbar/p_{\rm at}$, whereas the atomic unit of electric field strength
equals ${\cal E}_{\rm at}=\alpha^2m_{\rm e}c^2/(|e|a_0)$.

\section{Theoretical formulation}
\label{sec:theory}
The theoretical derivations presented here are based on the SFA. As it was shown in Refs.~\cite{Circular,VonKarmaan,Trains}, 
the SFA is an excellent analytical tool for the treatment of photodetachment from negative ions. A direct comparison between 
the results obtained within this framework and the numerical solution of the time-dependent Schr\"odinger equation has shown 
remarkable similarities. This is actually expected, as the SFA neglects the Coulomb interaction between the freed electron 
and the parent ion. In laser-induced photodetachment the residue is a neutral atom, hence the Coulomb interaction during the electron 
evolution in the continuum is absent. In contrast, in photoionization, the parent ion has a positive charge and the Coulomb potential 
modifies the electron dynamics, particularly in the low-energy regime. However, the SFA presents several advantages as compared to other 
more sophisticated ab initio calculations, including a faster numerical computation and simpler analytical expressions, which can be used 
to understand, in a deeper way, the physical phenomena under consideration.

For the reasons stated above, we shall limit our analysis of photodetachment from negative ions to the framework of the SFA. Even though 
the probability amplitude of detachment was initially calculated by Gribakin and Kuchiev in Ref.~\cite{Gribakin1997}, and further explored 
elsewhere (see, e.g.,~\cite{GroundState,Trains}), here we shall present the most important results along its derivation.

\subsection{Probability amplitude of photodetachment}
\label{amplitude}

It is assumed that, in the remote past, the electron is found in the ground state of the H$^-$ anion ($s$-electron) of energy $E_0$, which we 
denote as $\Phi_{0}({\bm r},t)=\ee^{-\ii E_0 t}\Phi_{0}({\bm r})$. By the action of the laser field, which lasts for a time $T_p$, 
the electron is promoted to the continuum. The probability amplitude of photodetachment, under the scope of the SFA, is given by~\cite{Gribakin1997,Circular,VonKarmaan,GroundState,Trains}
\begin{equation}
{\mathcal A}({\bm p})=-\ii \int_{0}^{T_p} \dd t\int\dd^3 r\,\psi^*_{\bm p}({\bm r},t)H_{\rm I}({\bm r},t)\Phi_0({\bm r},t),
\label{SFA1}
\end{equation}
where $\psi_{\bm p}({\bm r},t)$ is the Volkov state of the electron in the laser field~\cite{Wolkow1935} with an asymptotic momentum 
${\bm p}$. In the following, we shall consider only the length gauge for our calculations, as it was suggested in Ref.~\cite{Gribakin1997}. The Volkov solution, in this gauge, reads
\begin{equation}
\psi_{\bm p}({\bm r},t)=\exp\Big[\ii [{\bm p}-e{\bm A}(t)]\cdot{\bm r}-\frac{\ii}{2\me}\int_0^t\dd t'({\bm p}-e{\bm A}(t'))^2\Big],
\label{SFA2}
\end{equation}
with ${\bm A}(t)$ defining the vector potential of the laser field in the dipole approximation. The interaction Hamiltonian, $H_{\rm I}({\bm r},t)$, takes the form
\begin{equation}
{H}_{\rm I}({\bm r},t)=-e\bm{{\mathcal{E}}}(t)\cdot{\bm r},
\label{SFA3}
\end{equation}
where $\bm{{\mathcal{E}}}(t)=-\partial_t{\bm A}(t)$ is the oscillating electric field corresponding to the vector potential ${\bm A}(t)$. 

The unperturbed ground-state wave function of the electron in the H$^-$ anion, $\Phi_0({\bm r})$, is determined according to the zero-range potential model~\cite{Gribakin1997,Smirnov}
\begin{equation}
\Phi_{0}({\bm{r}})= \frac{A}{\sqrt{4\pi}a_0}\frac{\ee^{-\kappa r/a_0}}{r},
\label{GS}
\end{equation}
where $A$ is a scaling factor chosen such that the properties of the anion coincide with more advanced ab initio calculations or experimental data~\cite{Smirnov}, and $\kappa$ relates to the ground state energy, $E_{0}=-(\alpha c)^2m_{e}\kappa^2/2$. For our numerical illustrations we use the values $A=0.75$ and $\kappa=0.2354$~\cite{Gribakin1997}, which correspond to an ionization potential for H$^-$ equal to $|E_0|\approx0.754$~eV.

From Eqs.~\eqref{SFA1} to~\eqref{SFA3}, and after some algebraic manipulations (for details see, e.g., Refs.~\cite{GroundState,Trains}), the probability amplitude of detachment takes the form
\begin{equation}
{\mathcal A}({\bm p})=\ii e\int_0^{T_p}\dd t\,{\bm{\mathcal E}}(t)\cdot \tilde{\bm \Phi}_0({\bm p}-e{\bm A}(t))\ee^{\ii G_{\bm p}(t)}.
\label{SFA4}
\end{equation}
Here, we have introduced the Fourier transform of the ground-state wave function, 
\begin{equation}
\tilde{\bm{\Phi}}_{0}(\bm{p})=\ii\nabla_{p}\tilde{\Phi}_{0}(\bm{p})=-\ii\frac{4\sqrt{\pi}A}{(\kappa^{2}+\bm{p}^2)^{2}}{\bm{p}},
\label{SFA5}
\end{equation}
and the function $G_{\bm p}(t)$ defined as
\begin{equation}
G_{\bm p}(t)=\frac{1}{2\me}\int_0^t\dd t'\Big[{\bm p}-e{\bm A}(t')\Big]^2-E_0t.
\label{SFA6}
\end{equation}
Hence, by integrating numerically Eq.~\eqref{SFA4} and taking into account Eqs.~\eqref{SFA5} and~\eqref{SFA6}, we calculate the probability 
amplitude of photodetachment. The driving laser field used in our illustrations is described in Sec.~\ref{field}, after we comment on 
the properties of ${\mathcal A}({\bm p})$ in the three-dimensional (3D) momentum space.

\subsection{Vortices vs Nodes}

The differences between nodal surfaces and vortex lines have been thoroughly analyzed in Ref.~\cite{Circular}. However, for the sake of 
clarity, we present here their main properties. As presented in Ref.~\cite{Bialynicki2000}, vortex lines appear in the 3D momentum space as closed loops or continuous curves
(starting and ending at infinity). While the complex probability amplitude of detachment ${\cal A}({\bm p})$ vanishes
along those lines, around them the amplitude's phase changes from $0$ to $2\pi m$, where $m=\pm1,\pm2,...$ is an integer number 
called the {\it topological charge} (see, e.g., Refs.~\cite{Lloyd2017,Bliokh2017}). In contrast, nodes appear as surfaces of vanishing 
probability, accross which the phase of ${\cal A}({\bm p})$ jumps by $\pm\pi$. Note that in a two-dimensional plane, known as Poincar\'e section, 
vortex lines are typically visualized as single points, while nodal surfaces appear as curves of zero probability. 

As was elaborated in Ref.~\cite{Circular}, the appearance of vortices in photodetachment can be quantified by the quantization condition,
\begin{equation}
m=\frac{1}{2\pi}\oint_{\cal C} {\bm v}({\bm p})\cdot \dd{\bm p},
\label{m1}
\end{equation}
where the velocity field ${\bm v}({\bm p})$ is related to the probability amplitude of photodetachment ${\mathcal A}({\bm p})$,
\begin{equation}
{\bm v}({\bm p})=\frac{1}{|{\mathcal A}({\bm p})|^2} {\rm Re}\,[{\mathcal A}^*({\bm p})(-\ii{\bm \nabla}_{\bm p}){\mathcal A}({\bm p})].
\label{m2}
\end{equation}
For simplicity, we assume that the Poincar\'e section, which contains the closed path of integration ${\cal C}$ in Eq.~\eqref{m1}, coincides with the $p_xp_y$-plane 
(i.e., we set $p_z=0$). More precisely, we choose ${\cal C}$ to be a circular contour of radius $p_r$ centered at momentum ${\bm p}_0=(p_{0x},p_{0y},0)$.
Hence, it can be parametrized by an angle $\varphi_{\bm p}$ such that
\begin{equation}
p_x=p_{0x}+p_r\cos{\varphi_{\bm p}},\; p_y=p_{0y}+p_r\sin{\varphi_{\bm p}}, \; p_z=0.
\label{NL1}
\end{equation} 
It is straighforward to show in this case that the quantization condition~\eqref{m1} becomes
\begin{equation}
m(p_r)=\frac{1}{2\pi}\int_0^{2\pi}\dd\varphi_{\bm p}\frac{{\rm Im}\big[{\mathcal A}^*({\bm p})\partial_{\varphi_{\bm p}}{\mathcal A}({\bm p})\big]}{|{\mathcal A}({\bm p})|^2},
\label{NL3}
\end{equation}
where $m=m(p_r)$. Actually, the contour ${\cal C}$ may contain multiple vortices. In such case, $m(p_r)$ represents a total topological charge
enclosed by the contour. If, however, ${\cal C}$ encircles an isolated vortex point, which for instance can be realized if an
individual vortex stays enclosed by the contour in the limit $p_r\rightarrow 0$, Eq.~\eqref{NL3} represents its individual topological charge.
This definition will be used in Secs.~\ref{sec:CEP} and~\ref{sec:control} to interpret our numerical results. 

\subsection{Laser field}
\label{field}

\begin{figure*}
\centering
\includegraphics[width=18cm]{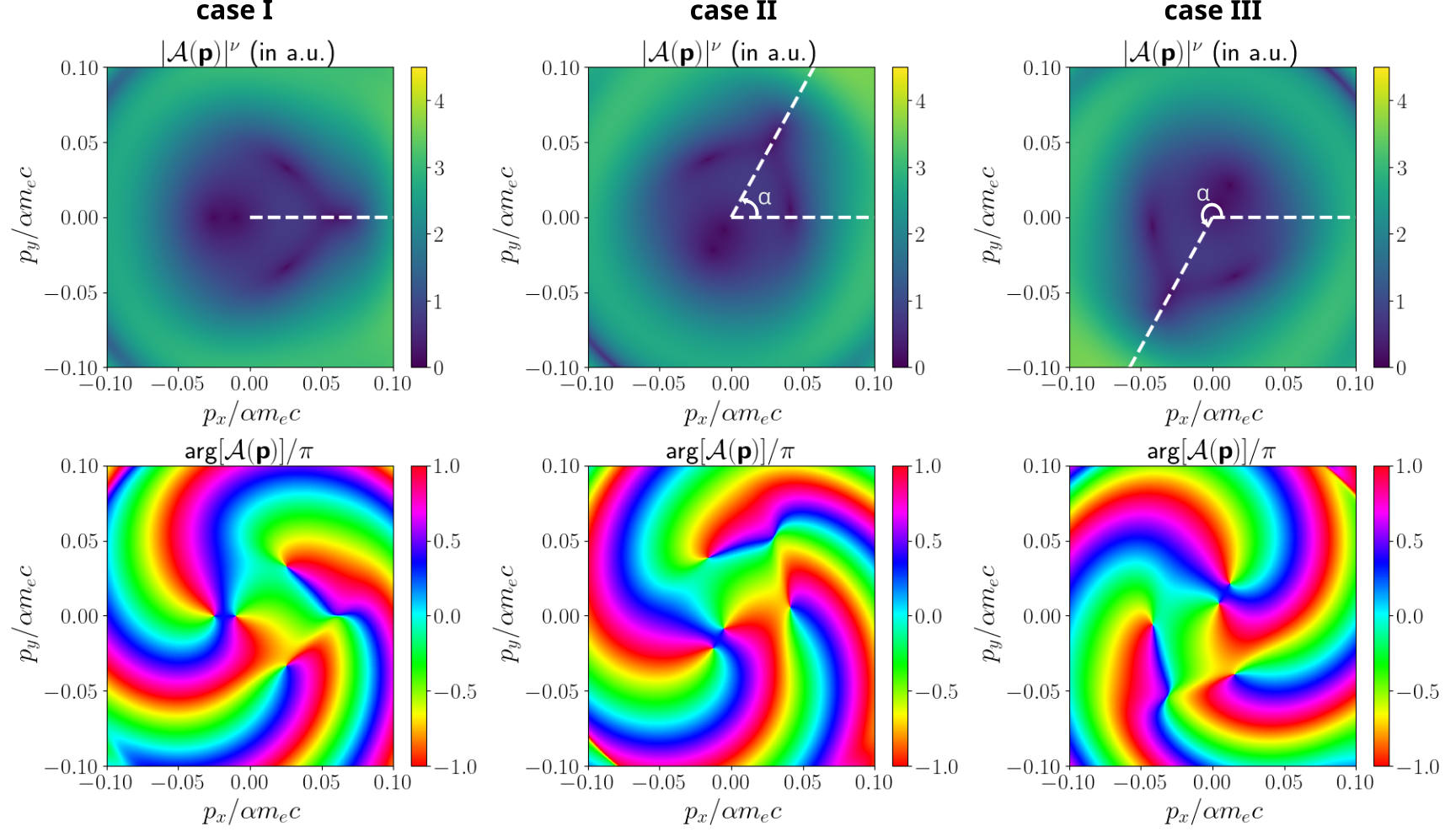}
\caption{Modulus (upper panels) and phase (lower panels) of the probability amplitude of photodetachment ${\mathcal A}({\bm p})$ 
[Eqs.~\eqref{SFA4} to~\eqref{SFA6}] in the $p_xp_y$-plane ($p_z=0$). The modulus has been raised to the power $\nu=0.5$ and the phase 
is presented in units of $\pi$, which is done only for visual purposes. A circularly-polarized driving pulse  ($N_{\rm rep}=1$) 
comprising $N_{\rm osc}=3$ oscillations within a $\sin^2$ envelope [see, Eqs.~\eqref{pulse1} to~\eqref{pulse3}] has been used. 
While we keep its peak intensity ($I=2.5\times 10^{11}$W/cm$^2$) and wavelength ($\lambda=4000$~nm) fixed, each column corresponds to
a different CEP and helicity of the pulse. Specifically, the left column (Case I) is for the laser field configuration 
$(\chi,\sigma)=(0,+)$, meaning that the CEP is zero and the laser pulse rotates counterclockwise. The middle column (Case II) 
corresponds to the configuration $(\pi/3,+)$, whereas the right column (Case III) corresponds to the configuration $(2\pi/3,-)$.}
\label{rys3}
\end{figure*}

As we plan to demonstrate in this paper, the properties of the laser field are imprinted onto vortical structures of photodetachment. For this reason,
a careful definition of the driving laser field is important. We define an isolated $N_{\rm osc}$ cycle laser pulse, lasting for time 
$\tau_p$, by the following time-dependent electric field, 
\begin{equation}
\bm{{\mathcal{E}}}_{\sigma\chi}(t)=\begin{cases}
{\cal E}_0\sin^2\bigl(\frac{\omega t}{2N_{\rm osc}}\bigr)\bm{F}(t,\sigma,\chi), & 0\leqslant t\leqslant \tau_p , \cr
0, & \textrm{otherwise},
\end{cases}
\label{pulse1}
\end{equation}
with the circularly-polarized carrier wave,
\begin{equation}
\bm{F}(t,\sigma,\chi)=\sin(\omega t+\chi){\bm e}_x-\sigma\cos(\omega t+\chi){\bm e}_y,
\label{pulse2}
\end{equation}
and ${\cal E}_0={\cal E}_{\rm at}\sqrt{\frac{I}{2I_{\rm at}}}$.
Here, $I$ denotes the peak intensity of the pulse expressed in units of $I_{\rm at}=3.51\times 10^{16}$~W/cm$^2$, $\omega=2\pi N_{\rm osc}/\tau_p$ is 
the carrier frequency of the field, whereas $\chi$ is the carrier envelope phase of the pulse. 
The polarization properties of the field are controlled by the helicity, $\sigma$, and in our further analysis we shall choose circularly-polarized pulses
with either counterclockwise ($\sigma=+1$) or clockwise ($\sigma=-1$) orientations. We also define a train of $N_{\rm rep}$ such pulses,
\begin{equation}
\bm{{\mathcal{E}}}_{\sigma_1\chi_1,\sigma_2\chi_2,\ldots}(t)=\sum_{\ell=1}^{N_{\rm rep}} 
\bm{{\mathcal{E}}}_{\sigma_\ell\chi_\ell}\Bigl(t-(\ell-1)\tau_p\Bigr),
\label{pulse3}
\end{equation}
with, in principle, different $\sigma_\ell$ and $\chi_\ell$. (Note that, under such conditions, the total duration of the train of pulses is $T_p=N_{\rm rep}\tau_p$.) 
Hence, the electromagnetic vector potential describing the laser field can be defined,
\begin{equation}
{\bm A}_{\sigma_1\chi_1,\sigma_2\chi_2,\ldots}(t)=-\int_{-\infty}^t\bm{{\mathcal{E}}}_{\sigma_1\chi_1,\sigma_2\chi_2,\ldots}(t')\dd t',
\label{pulse4}
\end{equation}
where it holds that
\begin{equation}
\lim_{t\rightarrow \pm\infty} \bm{A}_{\sigma_1\chi_1,\sigma_2\chi_2,\ldots}(t) =\bm{0}.
\label{pulse5}
\end{equation}
While those definitions are very general, in the following we shall consider either an isolated pulse ($N_{\rm rep}=1$) or a sequence of two pulses 
($N_{\rm rep}=2$) with corotating or counterrotating circular polarizations. For our numerical illustrations we shall keep the peak intensity, 
wavelength, and number of field oscillations within the $\sin^2$ envelope fixed ($I=2.5\times 10^{11}$W/cm$^2$, $\lambda=4000$~nm, and 
$N_{\rm osc}=3$, respectively), while allowing the CEP and helicity to vary independently for each pulse in the train. Hence, we use the 
notation $(\chi_1,\sigma_1;...;\chi_\ell,\sigma_\ell;...;\chi_{N_{\rm rep}},\sigma_{N_{\rm rep}})$ to identify each field configuration. From $\sigma_\ell$ we only retain its sign, i.e., we write $\sigma_\ell=\pm$.

\section{Laser-field control of vortical structures}
\label{sec:CEP}
In this Section, we reveal the relationship between the CEP of the driving laser field and the formation of electron vortices in photodetachment. 
As mentioned above, we fix the number of field oscillations within each individual pulse, its peak intensity, and wavelength, while allowing to vary the number of pulses in the train, the individual CEPs, and helicities. Our main objective is to determine how such parameters influence the formation of electron vortices in photodetachment and the overall probability distribution of photoelectrons.

\subsection{Isolated laser pulse}
We start by analyzing the creation of electron vortices in photodetachment driven by a single laser pulse  
[$N_{\rm rep}=1$ in Eq.~\eqref{pulse3}]. In Fig.~\ref{rys3}, we present the color mappings of detachment probability amplitude   
${\mathcal A}({\bm p})$ [Eqs.~\eqref{SFA4} to~\eqref{SFA6}] calculated in the $p_xp_y$-plane ($p_z=0)$. While in the upper panels we show 
the amplitude's modulus, i.e.,  $|{\mathcal A}({\bm p})|^\nu$, where $\nu=0.5$ has been chosen for visual purposes, in the lower panels 
we present its phase,  ${\rm arg}[{\mathcal A}({\bm p})]/\pi$. We have considered pulses of different CEPs, labelled as Case I (left column), 
Case II (middle column), and Case III (right column), which correspond to the configurations $(0,+)$, $(\pi/3,+)$, and $(2\pi/3,-)$, 
respectively.

\begin{figure*}
\centering
\includegraphics[width=18cm]{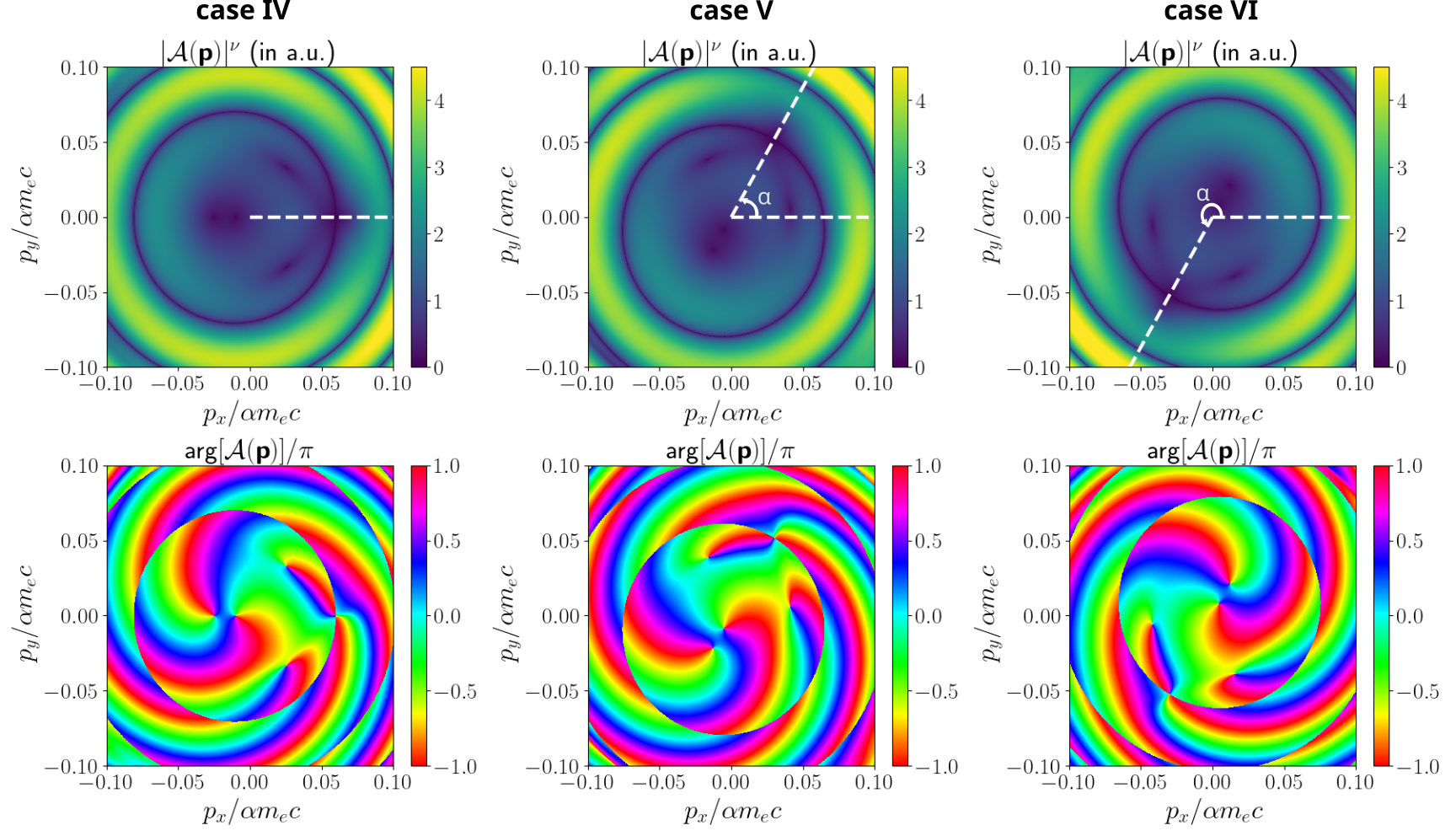}
\caption{The same as in Fig.~\ref{rys3} but for a pair of two identical laser pulses ($N_{\rm rep}=2$)
[configuration $(\chi,\sigma;\chi,\sigma)$]. The left column (Case ${\rm IV}$) is for the configuration 
$(0,+;0,+)$ (see, Ref.~\cite{Circular}), the middle column (Case V) is for $(\pi/3,+;\pi/3,+)$, whereas the right column 
is for $(2\pi/3,-;2\pi/3,-)$. }
\label{rys4}
\end{figure*}

In Case I (left column of Fig.~\ref{rys3}), we observe the appearance of four vortices; they manifest themselves as points of vanishing 
probability for which the amplitude's phase changes continuously  from $-\pi$ to $\pi$ around them. Hence, we assign to each of them the topological 
charge $m=1$. Two of such vortices are located along 
the line $p_y=0$ while the other two appear at $p_x=0.025\alpha\me c$~\cite{Comment2}. In the middle panels (Case II), which correspond to 
a driving field of the type $(\pi/3,+)$, we observe the exact same configuration of vortices, but rotated by an angle $\alpha=\pi/3$ with 
respect to the center of coordinates. Finally, in the right column of Fig.~\ref{rys3} (Case III), we present the results corresponding 
to the driving pulse configuration $(2\pi/3,-)$, such that the CEP takes now the value $2\pi/3$ and the helicity is inverted, $\sigma=-1$ 
(clockwise field evolution). This time, we deal with antivortices ($m=-1$) and the rotation angle $\alpha=4\pi/3=-2\pi/3$. Hence, we conclude that the 
effect of the CEP and helicity of the driving pulse ($N_{\rm rep}=1$) over the vortex pattern (or, in general, over the probability 
distribution of photoelectrons in momentum space) is an overall rotation by an angle $\alpha=\chi\sigma$ with respect to the origin of 
coordinates. 

\subsection{Two corotating laser pulses }
Up to now we have considered the vortex formation stimulated by isolated laser pulses. However, it is interesting to analyze 
the case when a pair of pulses [$N_{\rm rep}=2$ in Eq.~\eqref{pulse3}] interacts with the H$^-$ ion. We shall start by exploring the photodetachment driven by such a train in the corotating configuration ($\sigma_1=\sigma_2\equiv\sigma$) with identical CEPs ($\chi_1=\chi_2\equiv\chi$).

In Fig.~\ref{rys4} we show the same as in Fig.~\ref{rys3} but this time the driving field consists of two corotating pulses ($N_{\rm rep}=2$)
and we vary the CEP, which is common for both pulses. According to our convention,  
the driving fields considered here are labelled as $(\chi,\sigma;\chi,\sigma)$. In the left column of Fig.~\ref{rys4} (Case ${\rm IV}$), 
which corresponds to a configuration $(0,+;0,+)$, we observe the formation of the same four vortices as in  Case~I (see, the left column of 
Fig.~\ref{rys3}) with additional nodal surfaces. The latter appear as annular zones of zero probability located at different radii 
(see, Ref.~\cite{Circular}). Note that the regions of high probability resemble the Newton's rings reported in the perturbative 
photoionization of He atoms~\cite{Starace2015}. In the middle column of Fig.~\ref{rys4} (Case ${\rm V}$), which is for the configuration 
$(\pi/3,+;\pi/3,+)$ we see that the very same system of vortices and nodal surfaces is formed, but it is rotated by an angle 
$\alpha=\pi/3$ with respect to the origin of coordinates. Moreover, in the right column of Fig.~\ref{rys4} (Case VI) we present our 
results for the train configuration $(2\pi/3,-;2\pi/3,-)$, namely, both pulses are oriented clockwise, and both CEPs are equal to $2\pi/3$. 
Here we observe an identical pattern as in Case ${\rm IV}$ but rotated by an angle $\alpha=4\pi/3=-2\pi/3$~mod $2\pi$, while vortices
are transformed into antivortices. This indicates that the addition 
of a nonzero CEP, common to both pulses in the corotating train, leads to a rotation of the probability distribution of photoelectrons by 
an angle $\alpha=\sigma\chi$ in the polarization plane. The same has been seen in Fig.~\ref{rys3} for a single pulse. Note also that a similar dependence of the distribution of photoelectrons with the CEP 
has been observed in photoionization of molecules by trains of bichromatic laser pulses~\cite{Bandrauk2016}. 

A more interesting situation is met when one of the CEPs is kept constant (say $\chi_1$) and we allow the other one ($\chi$) to vary. 
In the Supplemental Material we show the magnitude of the probability amplitude (Animation 1) and its phase (Animation 2) for photodetachment 
driven by a train of two pulses ($N_{\rm rep}=2$) in the configurations $(\pi/3,+;\chi,+)$, where $\chi$ changes from $0$ to $2\pi$. 
From both magnitude and phase we can see that, at $\chi=0$, four well defined vortices with the same topological charge ($m=1$) 
and one antivortex ($m=-1$) are present. Also, an open ring of low probability is visible at larger photoelectron momenta. By changing 
the CEP there is a continuous expansion of the open ring, together with antivortex migration towards larger $|{\bm p}|$. When both 
antivortex and open ring merge, a full nodal surface (annular region of zero probability) is formed. This can be seen in Animation 1 
for $\chi\approx 0.9$~rad. Another interesting effect is the creation of vortex-antivortex pairs while increasing $\chi$ (see Animation 2), 
which leads to the formation of new open rings at smaller momentum. Hence, vortex-antivortex pair creation, their migration, and further 
recombination lead to a cyclic formation of nodal lines propagating outwards in the $p_xp_y$-plane while changing $\chi$. As expected, 
when $\chi=2\pi$, we end up with the same momentum pattern as for $\chi=0$. The formation of nodal surfaces (visualized as closed rings) 
suggests that the open annular zones propagate together with a single vortex. As a result, vortex-antivortex annihilation leads to the formation 
of the closed rings in the 2D space. Note that, in contrast to the results presented in 
Ref.~\cite{Starace2015} (perturbative regime), the probability amplitude of photodetachment in Animation~1 (nonperturbative regime) 
is not circularly symmetric. Hence, the formation of vortex patterns and the rotation of the high-probability structures can be observed 
with increasing the relative CEP.

\subsection{Two counterrotating laser pulses}
We have seen that single driving pulses and pairs of two corotationg pulses with the same CEP lead to vortex structures and nodal surfaces 
in the probability amplitude of detachment. A change in the CEP (or helicity) has the effect of a rotation of the probability amplitude 
by an angle $\alpha=\sigma\chi$ with respect to the origin of coordinates. In this Section, we shall analyze the electron photodetachment driven 
by a pair of counterrotating pulses ($\sigma_1=1$ and $\sigma_2=-1$). First we assume a common CEP, i.e., we set $\chi_1=\chi_2\equiv\chi$, 
meaning that the laser field configurations are of the type $(\chi,+;\chi,-)$. In the Supplemental Material we show the modulus of the 
probability amplitude of detachment (Animation 3) and its phase (Animation 4) with $\chi$ changing from $0$ to $\pi$. This time, the effects 
of the CEP are much more complex than a simple rotation of momentum distribution in the $p_xp_y$-plane. For $\chi=0$, we observe the presence of momentum spirals 
(see, Ref.~\cite{Circular}) with no vortices. However, a continuous formation and annihilation of 
vortex-antivortex pairs with topological charges $m=1$ and $m=-1$, respectively, occurs for increasing $\chi$. Finally, at $\chi=\pi$ 
the original momentum spiral is recovered, but rotated by an angle $\pi$ in the polarization plane. Hence, vortical structures are observed for all CEPs, except for the particular cases $\chi=0$ or $\chi=\pi$.

\begin{figure*}
\centering
\includegraphics[width=18cm]{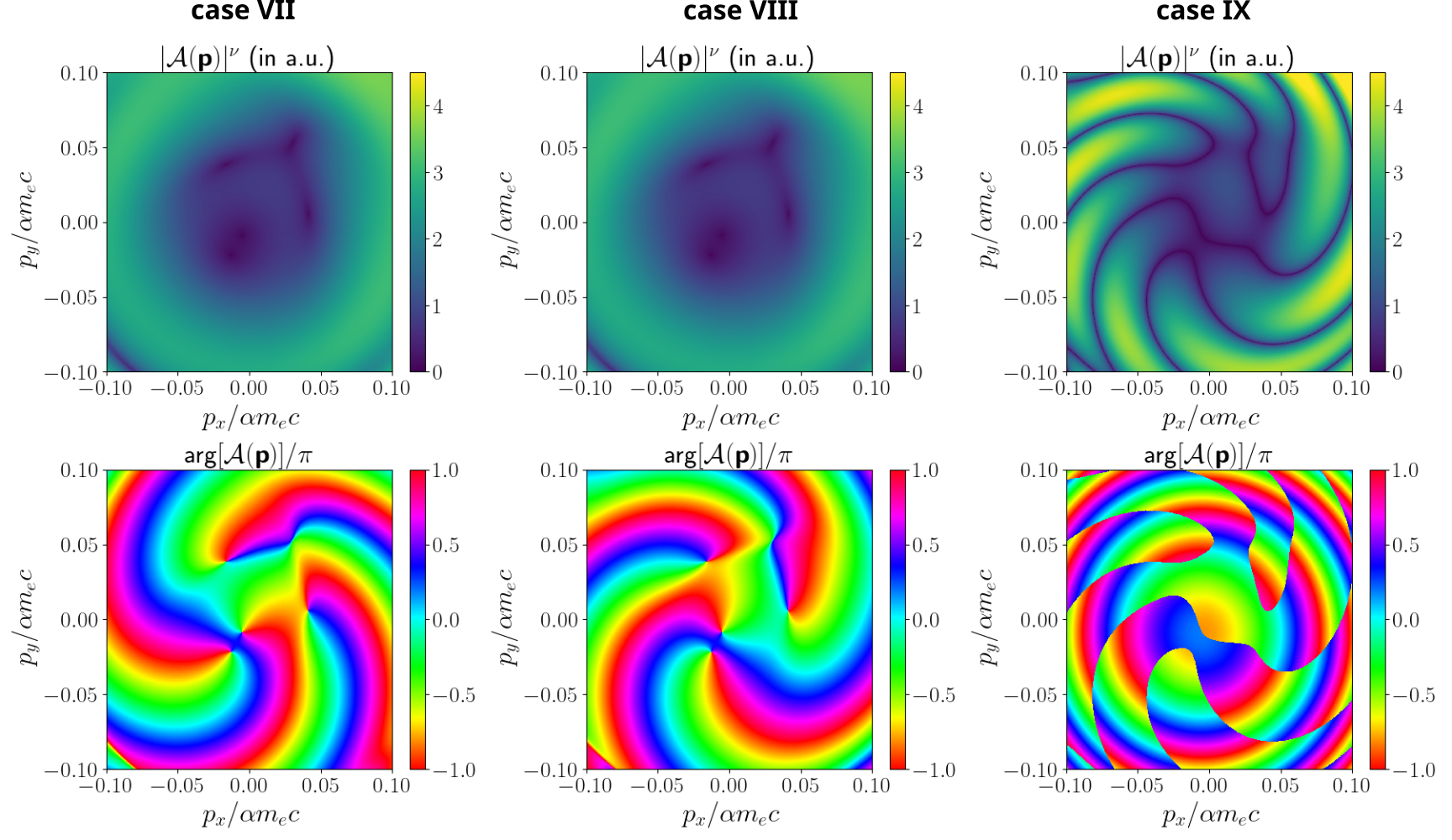}
\caption{The same as in Figs.~\ref{rys3} and~\ref{rys4} but for the following laser field configurations: the left column (Case VII) 
corresponds to photodetachment driven by a single laser pulse of the type $(\pi/3,+)$; the middle column (Case VIII) is obtained from 
a single pulse in the configuration $(-\pi/3,-)$; and the right column (Case IX) corresponds to photodetachment driven by a pair of such 
pulses in the configuration $(\pi/3,+;-\pi/3,-)$. }
\label{rys5}
\end{figure*}

In Animations 5 and 6 of the Supplemental Materials we show the modulus and phase of the probability amplitude of detachment 
${\mathcal A}({\bm p})$, respectively, for laser fields in the configuration $(\chi_1,+;\chi,-)$, with $\chi_1=\pi/3$. 
While the CEP of the first pulse is fixed, 
the CEP of the second pulse, $\chi$, varies from $0$ to $2\pi$. At $\chi=0$, both spirals and vortex-antivortex pairs are observed 
within the region of momentum considered here. With increasing $\chi$, the spirals start to rotate and vortex-antivortex pairs are continuously 
created or annihilated. It is worth mentioning that a CEP $\chi=-\pi/3=5\pi/3\approx 5.2$~rad leads to the total annihilation 
of votex-antivortex pairs and only the spiral remains. The disappearance of vortical structures occurs when $\chi=-\chi_1$, which is
directly related to the time-reversal properties of the laser field. We note that a train of pulses in the configuration $(0,+;0,-)$ also 
leads exclusively to momentum spirals (no vortices) which are rotated by an angle $-\chi_1=-\pi/3$  with respect to the configuration 
$(\pi/3,+;-\pi/3,-)$  (see, the right panels of Fig.~4 in Ref.~\cite{Circular}).  The time-reversal symmetry leading to mutual 
annihilation of vortex-antivortex pairs, and the observation of spirals, will be discussed next.

\section{Vortex-antivortex annihilation and time reversibility}
\label{sec:control}

In this Section, we focus on the case when detachment is driven by a pair of counterrotating circularly-polarized laser pulses with opposite 
CEPs [i.e., the laser fields are in the configurations $(\chi,\sigma)$ and $(-\chi,-\sigma)$ for single pulses or $(\chi,\sigma;-\chi,-\sigma)$
for trains of two pulses]. As an example, in Fig.~\ref{rys5} we plot the color mappings of probability amplitude of ionization ${\cal A}({\bm p})$ for: an isolated pulse 
with $\sigma=1$ and $\chi=\pi/3$ (left column, Case VII), the same with $\sigma=-1$ and $\chi=-\pi/3$ (middle column, Case VIII), and finally the
sequence of those pulses (right column, Case IX). For the latter, we obtain the exactly same pattern irrespectively of the order of pulses in a sequence.
As before, we keep $p_z=0$, and we present separately the modulus and the phase of ${\cal A}({\bm p})$ in each case.

When comparing the results for individual pulses (left and middle columns of Fig.~\ref{rys5}), the absolute values of probability amplitudes 
turn out to be identical. Specifically, the zeroes of ${\cal A}(\bm p)$ in both cases are located at the exactly same momenta. As it follows
from the mappings of the phase of probability amplitudes, those points have a vortex character. More specifically, in both cases we observe two pairs 
of vortex points, with $m=1$ in Case VII and $m=-1$ in Case VIII.
Those properties can be explained if one realizes that the respective probability amplitudes, in Cases VII and VIII, 
are related by the time-reversal transformation: $t\longmapsto -t$. Namely, it follows from Eq.~\eqref{pulse2}
that $\bm{F}(t,-\sigma,-\chi)=-\bm{F}(-t,\sigma,\chi)$ and, hence, 
$\bm{{\mathcal{E}}}_{-\sigma-\chi}(t)\longmapsto-\bm{{\mathcal{E}}}_{\sigma\chi}(t-\tau_p)$. Taking into account that under the time-reversal symmetry 
the wave functions are transformed into their complex conjugates [i.e., $\psi_{\bm p}({\bm r},t)\longmapsto \psi_{\bm p}^*({\bm r},t)$
and $\Phi_0({\bm r},t)\longmapsto \Phi_0^*({\bm r},t)$ in Eq.~\eqref{SFA1}], one concludes that the probability amplitudes of ionization driven by pulses
of opposite helicities and CEPs are related such that
\begin{equation}
{\cal A}({\bm p};-\sigma,-\chi)=[{\cal A}({\bm p};\sigma,\chi)]^*\,\ee^{-\ii\tau_p({\bm p}^2/2m_{\rm e}-E_0)}.
\label{tr1}
\end{equation}
Here, we have explicitly distinguished between the probability amplitudes calculated for those pulses. 
It follows from Eq.~\eqref{tr1} that $|{\cal A}({\bm p};-\sigma,-\chi)|=|{\cal A}({\bm p};\sigma,\chi)|$, whereas
${\rm arg}[{\cal A}({\bm p};-\sigma,-\chi)]=-{\rm arg}[{\cal A}({\bm p};\sigma,\chi)-\tau_p({\bm p}^2/2m_{\rm e}-E_0)$.
Note that an extra phase term here does not contribute to the contour integral in Eq.~\eqref{NL3}. 
%For a pulse of chirality $\sigma$ and CEP $\chi$, this condition can be written explicitely as
%%If, therefore, this condition 
%%defines the vortex of strength $m$ for the given pulse of chirality $\sigma$ and CEP $\chi$,
%\begin{equation}
%\oint_K {\bm \nabla}{\rm arg}[{\cal A}({\bm p};\sigma,\chi)]\cdot \dd{\bm p}=2\pi m, \quad m\in{\mathbb{Z}},
%\end{equation}
%Therefore, for the pulse of an opposite chirality and CEP, also the topological charge will change the sign. 
This explains why the vortex-like points, while located at the same momenta, will carry opposite topological charges in both cases.
%Now, representing ${\cal A}({\bm p};\sigma,\chi)=|{\cal A}({\bm p};\sigma,\chi)|\ee^{\ii\phi({\bm p})}$, we realize from Eq.~\eqref{} 
%that the phase $\phi({\bm p})$ around each isolated vortex (antivortex) satisfies the quantization condition,

A very distinct spiral pattern is observed when ionization occurs in a sequence of both pulses (right column in Fig.~\ref{rys5}, Case IX). In this case,
there is no vortices. Instead, we observe the nodal lines of the probability amplitude, with its phase jumping by $\pm\pi$ accross those lines.
These features can be explained by the formula, 
\begin{eqnarray}
{\cal A}({\bm p})&=&{\cal A}({\bm p};\sigma,\chi)+{\cal A}({\bm p};-\sigma,-\chi)\nonumber\\
&=&2|{\cal A}({\bm p};\sigma,\chi)|\ee^{-\frac{\ii\tau_p}{2}({\bm p}^2/2m_{\rm e}-E_0)}\nonumber\\
&\times&\cos\Bigl[{\rm arg}[{\cal A}({\bm p};\sigma,\chi)]+\frac{\tau_p}{2}\Bigl(\frac{{\bm p}^2}{2m_{\rm e}}-E_0\Bigr)\Bigr],
\label{tr2}
\end{eqnarray}
which defines the probability amplitude of ionization by a sequence of time-reversed pulses. It shows that the previously observed vortices 
and antivortices, located at the same momenta but with opposite topological charges, annihilate each other into simple nodes. At the same time, 
new nodes appear as indicated by the cosine function in Eq.~\eqref{tr2}. Since ${\rm arg}[{\cal A}({\bm p};\sigma,\chi)]$ changes by 
$2\pi$ (or by $-2\pi$) around an old vortex (antivortex), the cosine function has only two zeroes in their close vicinity at the fixed ${\bm p}^2$.
With increasing ${\bm p}^2$, the nodal curve is formed. The cosine function also contributes to the phase of the total probability amplitude in Eq.~\eqref{tr2}.
Namely, it changes the sign while passing across the node, which results in a disconuity of ${\rm arg}[{\cal A}({\bm p};\sigma,\chi)]$ by 
$\pm\pi$ across the nodal line. This explains how the momentum spirals in photodetachment are formed.

\section{Conclusions}
\label{sec:conclusions}
We have presented an in-depth analysis of the formation of electron vortices and momentum spirals in photodetachment driven by either an
isolated laser pulse or a pair of pulses with circular polarization. We have seen that, by changing the CEP and helicity of the individual 
pulses, it is possible to control the vortical structures and the appearance of spirals in the momentum distributions of photoelectrons.

We have shown that, in the presence of a circularly-polarized laser pulse or two corotating pulses, the probability distribution 
of photoelectrons exhibits only vortices or circular nodal lines (no spirals are observed). In addition, a rotation of such vortex structures 
by an angle $\alpha$ is consistent with the introduction of a nonvanishing CEP  [configurations $(\chi,\sigma)$ and $(\chi,\sigma;\chi,\sigma)$ 
with $\chi\neq0$] such that $\alpha=\sigma\chi$. This, in principle, can find applications in ultrashort laser pulse diagnostics, as the CEP 
can be directly inferred from the vortex pattern orientation in the two-dimensional momentum distribution of photoelectrons. Hence, it is 
expected that the CEP can be experimentally determined with great accuracy by a careful measurement of the rotation angle $\alpha$. 

When the driving field consists of two corotating pulses with different CEPs [configuration $(\chi_1,\sigma;\chi,\sigma)$ where $\chi_1$ 
is kept unchanged], the formation of vortices and antivortices is observed. Moreover, an increasing $\chi$ leads to a cyclic outwards propagation 
of nodal rings in the probability distribution of photoelectrons.

We have observed that spirals do only appear in the case of counterrotating laser pulses, which is in agreement with Ref.~\cite{Circular}. 
More generally, the driving field configurations of the type 
$(\chi,+;\chi,-)$ lead to generation of spirals and vortex-antivortex pairs. The latter are created and annihilated with increasing $\chi$. 
However, at $\chi=0$ or $\chi=\pi$ the distributions of photoelectrons exhibit only spirals, and they are rotated by an angle $\pi$ with 
respect to one another. 

Finally, the laser fields labelled as $(\chi_1,+;\chi,-)$, with $\chi_1$ constant, lead to rotating spirals while increasing $\chi,$ 
together with vortex-antivortex generation. Again, only spirals are observed when $\chi=-\chi_1$, which is a direct consequence of the 
time-reversal symmetry of the pulse train.

We have shown analytically that spectra of photoelectrons containing only spirals are formed when the driving field comprises two pulses 
with time-reversal symmetry. In such case, the vortex-antivortex patterns superimpose each other leading to spirals. 
We have also demonstrated that the phase of ${\cal A}({\bm p})$ changes by $\pm\pi$ while crossing the spiral arms.

\section*{Acknowledgements}
This work is supported by the National Science Centre (Poland) under Grant 
Nos. 2018/31/B/ST2/01251 (F.C.V.) and 2018/30/Q/ST2/00236 (J.Z.K. and K.K.).


\begin{thebibliography}{99}
\bibitem{Comment2}
The results shown in the left panels of Fig.~\ref{rys3} (Case I) were already presented in Fig. 4 of Ref.~\cite{Circular}. However, we show them here for two main reasons: firstly, by comparing our results with those presented in Ref.~\cite{Circular}, we corroborate that the computational code used in our calculations leads to accurate results. Secondly, the case $(0,+)$ is a benchmark to determine the influence of both CEP and helicity over vortex formation.
\end{thebibliography}
\end{document}